# Expurgated PPM Using Symmetric Balanced Incomplete Block Designs

Mohammad Noshad, *Student Member, IEEE,* and Maïté Brandt-Pearce, *Senior Member, IEEE*

*Abstract*—In this letter, we propose a new pulse position modulation (PPM) scheme, called expurgated PPM (EPPM), for application in peak power limited communication systems, such as impulse radio (IR) ultra wide band (UWB) systems and free space optical (FSO) communications. Using the proposed scheme, the constellation size and the bit-rate can be increased significantly in these systems. The symbols are obtained using symmetric balanced incomplete block designs (BIBD), forming a set of pair-wise equidistance symbols. The performance of $Q$-ary EPPM is better than any $Q$-ary pulse position-based modulation scheme with the same symbol length. Since the code is cyclic, the receiver for EPPM is simpler compared to multipulse PPM (MPPM).

## I. Introduction

M-ARY transmission has been a popular technique for increasing the data rate in communication systems. We present a spectrally efficient low-complexity $M$-ary modulation well suited to unipolar peak-power limited communication systems, as an alternative to the typical pulse position modulation (PPM) approach.

PPM is an $M$-ary technique that can be implemented non-coherently, and is therefore useful in impulse radio (IR) ultra wide band (UWB) radio frequency (RF) systems [1], and in free space optical (FSO) communications [2]. Although on-off keying (OOK) is the conventional binary modulation for FSO and UWB communications, PPM is preferred in systems that have weak intersymbol interference (ISI) effects. The advantage of PPM over OOK is evident in fading channels, since it does not need a threshold to make a decision on the received symbol.

In communication systems with peak power limited transmitters, the transmitted energy per symbol in PPM decreases with increasing symbol size, and this decreases the efficiency. Multipulse PPM (MPPM) has been proposed to solve this problem in FSO systems by transmitting pulses in multiple time-slots [3], [4]. This approach is also helpful in IR UWB systems, in which, for a fixed bit-rate, increasing the alphabet size reduces the average transmitted power [5]. But the complexity in encoding the symbols and mapping bits to symbols prevents MPPM from being widely used [6].

This research was funded by the National Science Foundation (NSF) under grant number ECCS-0901682.
Mohammad Noshad (mn2ne@virginia.edu) and Maïté Brandt-Pearce (mb-p@virginia.edu) are with Charles L. Brown Department of Electrical and Computer Engineering, University of Virginia Charlottesville, VA 22904.

Balanced incomplete block design (BIBD) is an area of combinatorial mathematics with many communication applications. Symmetric BIBDs are used as codewords in spectral-amplitude-coding (SAC) optical code division multiple access (OCDMA) networks [7]. Camtepe et. al. [8] use BIBDs for key distribution mechanisms in wireless sensor networks. A new construction technique for LDPC codes using special classes of BIBDs is proposed in [9]. In [10] BIBDs are used as symbols for $M$-ary transmission in ultraviolet (UV) non-line of sight (NLOS) links.

In this letter, a new PPM scheme called expurgated PPM (EPPM) using symmetric BIBDs is described. The symbols in this modulation technique are obtained by expurgating the symbols of multipulse PPM. We show that the minimum error probability is obtained when the symbols are equidistant from each other, which makes the bit-symbol mapping for the EPPM scheme as easy as for PPM and much simpler than for MPPM. We then show that by including the complements of the codewords the number of symbols can be doubled in EPPM, thereby improving the performance. Although BIBDs have been used in various applications, their optimality as an M-ary constellation has previously not been shown.

The rest of the letter is organized as follows. In Section II, we describe our new EPPM scheme. In Section III, we analyze the performance of the different PPM schemes. Numerical results are presented in Section IV. Section V concludes the letter.

## II. Description of Expurgated PPM Scheme

In PPM, the symbol time is divided into $Q$ equal time-slots, only one of which contains a pulse, forming a code with cardinality $Q$. In MPPM, every $Q$-sequence containing $K$ pulses is considered as a symbol, and so the code cardinality is increased to $\binom{Q}{K}$. Yet for a fixed $Q$ the minimum distance between the symbols in MPPM remains the same as for PPM, i.e., it does not increase by increasing $K$. For large $K$'s the MPPM code size becomes impractically large: the bit-symbol mappings become a problem both at the transmitter and receiver since the complexity grows with the number of symbols.

The idea in this paper is to choose $Q$ symbols out of all $Q$-tuples such that the resulting system has a minimum symbol error probability among all possible $Q$-tuple choices. We call the new scheme an expurgated PPM (EPPM), since the optimal $Q$-tuple set turns out to be a subset of the codewords from multipulse PPM.

Consider $Q$ symbols (binary sequences) of length $Q$ with weights $K_1, K_2, \ldots, K_Q$. Denote $\boldsymbol{C}$ as the $Q \times Q$ matrix formed using these symbols as rows. Let the sum of column $l$ of $\boldsymbol{C}$ be $W_l$, where

$$\sum_{l=1}^{Q} W_l = \sum_{l=1}^{Q} K_l = A. \quad (1)$$

The sum of all pair-wise distances can be expressed as

$$D = \sum_{i=1}^{Q}\sum_{j=1}^{Q} d_{ij} = 2(QA - \sum_{l=1}^{Q} W_l^2), \quad (2)$$

where $d_{ij}$ is the Hamming distance between symbols $i$ and $j$. Using the Cauchy Inequality we know that $\sum_{l=1}^{Q} W_l^2 \geq \frac{1}{Q}(\sum_{l=1}^{Q} W_l)^2$ and therefore

$$D \leq 2A(Q - \frac{A}{Q}). \quad (3)$$

We assume that the error probability between symbols $i$ and $j$ can be written in the form $f(d_{ij})$. Our approach is to minimize the union bound on the symbol error probability [1]

$$P_s^{(U)} = \frac{1}{Q}\sum_{i=1}^{Q}\sum_{\substack{j=1 \\ j \neq i}}^{Q} f(d_{ij}). \quad (4)$$

Let $f(.)$ be a convex and monotonically decreasing function, which is true for $\mathrm{erfc}(.)$, the complementary Gaussian error function. According to Jensen's inequality [11], $P_s^{(U)} \geq (Q-1)f(d)$, where $d = \frac{D}{Q(Q-1)}$ is the mean of the $d_{ij}$'s, and equality holds only if $d_{ij} = d$, for $i, j = 1, 2, \ldots, Q$ and $i \neq j$. According to this, the performance is optimized when symbols are pair-wise equidistant. Using (3), for fixed $A$ and $Q$, $P_s^{(U)}$ is minimized when $D = 2A(Q - \frac{A}{Q})$. Then, $d$ is maximized for $A = Q^2/2$ and its maximum value is $d_{\max} = \frac{Q^2}{2(Q-1)}$. For equidistant codes, $d$ should be an integer, and thus, its maximum is $\frac{Q+1}{2}$, which is $\frac{1}{2(Q+1)}$ smaller than $d_{\max}$; this cost of enforcing the equidistant property decreases as $Q$ increases.

For cyclic codes, each codeword is a cyclic shift of the other codewords, which makes the structure of the transmitter and receiver, and also the encoding of the symbols, simpler. Henceforth, we focus on cyclic codes, which have equal weight codewords, i.e., $K_i = K \,\forall i$. Consequently, the cross-correlation between each pair of symbols is constant and equal to $K - \frac{d}{2}$. This property leads us to symmetric BIBD.

In symmetric BIBD, a set $S$ with $Q$ elements (called points), $\{u_1, u_2, \ldots, u_Q\}$, is considered, i.e., $|S| = Q$ and $u_j \in S$. $Q$ subsets of $S$ (called blocks), denoted as $\{U_1, U_2, \ldots, U_Q\}$, each containing $K$ elements (points), are designed, i.e., $U_i \subset S$ and $|U_i| = K$, such that the number of common elements (points) in each pair of subsets (blocks) is $\lambda$, i.e., $|U_i \cap U_j| = \lambda$ for $i \neq j$ [12]. Symmetric BIBDs, denoted by $(Q, K, \lambda)$, have the following relation:

$$\lambda(Q-1) = K(K-1). \quad (5)$$

A $(Q, K, \lambda)$-design is determined by its incidence matrix:

[1] This is asymptotically optimal but may not provide the optimal solution for finite distances.

this is the binary $Q \times Q$ matrix $\boldsymbol{C} = [c_{ij}]$ defined by

$$c_{ij} = \begin{cases} 1 & \text{if } u_j \in U_i, \\ 0 & \text{otherwise} \end{cases}.$$

We propose to use a symmetric BIBD as a code and the rows of its incidence matrix as the codewords. Hence, for a design with parameters $(Q, K, \lambda)$, $Q$ is the number of codewords, which is equal to the code length, $K$ is the code weight, which is defined as the number of 1's in each codeword, and $\lambda$ is the cross-correlation between each pair of codewords. We denote codeword $j$ by $\boldsymbol{c}_j$, which is the $j$th row of matrix $\boldsymbol{C}$.

Among all available symmetric BIBDs with length $Q$, we want ones for which the Hamming distance between each pair of codewords is maximum. The distance between the codewords with parameters $(Q, K, \lambda)$ is $2(K-\lambda)$. Using (5), the distance is obtained as $d = 2K(\frac{Q-K}{Q-1})$. The optimum $K$ for maximizing $d$ is $Q/2$. In this work, we choose codes with $Q = 2K+1$ and $K = 2\lambda+1$, since they have distance closest to the maximum distance among all known cyclic codes. According to [12]- [13], cyclic BIBD codes are available when $Q = 2K+1$ and $K = 2\lambda+1$, for a vast range of $\lambda$. For these codes, the distance between the codewords is $(Q+1)/2$, and therefore, they are optimal equidistant codes. Fig. 1 shows the symbols for PPM and this new code, which we call EPPM, for $Q = 7$, $K = 3$ and $\lambda = 1$.

Adding the complement of the codewords as symbols can double the number of symbols. The Hamming distance between $\overline{\boldsymbol{c}}_j$, the complement of $\boldsymbol{c}_j$, and $\boldsymbol{c}_i$ is

$$d(\overline{\boldsymbol{c}}_j, \boldsymbol{c}_i) = \begin{cases} Q & ; i = j, \\ 2\lambda+1 & ; i \neq j \end{cases}.$$

So extending the code size decreases the minimum distance by one. We call this an augmented EPPM (AEPPM).

In our EPPM scheme, we use the $Q$ blocks of a cyclic BIBD as codewords. In this case, the optimal receiver assuming an AWGN channel can be implemented as a shift register followed by a differential circuit, as shown in Fig. 2 for $Q = 7$. This detector is equivalent to the correlation receiver. In this figure, $\boldsymbol{x} = \{x_1, x_2, \ldots, x_Q\}$ is the stored data received in $Q$ time-slots, and $\Gamma = \frac{\lambda}{K-\lambda}$. Symbol synchronization is a key issue for EPPM, as for PPM; similar synchronization algorithms can be used for either modulations. In each symbol period, the receiver generates $Q$ variables at the output of the differential circuit by circulating the stored data in the shift register. The combination of the shift register and the differential circuit generates the decision statistic $Z_j = \langle \boldsymbol{x}, \boldsymbol{c}_j \rangle - \Gamma \langle \boldsymbol{x}, \overline{\boldsymbol{c}_j} \rangle$, for $j = 1, 2, \ldots, Q$, where $\langle \boldsymbol{x}, \boldsymbol{y} \rangle$ denotes the dot product of the vectors $\boldsymbol{x}$ and $\boldsymbol{y}$. Due to the fixed cross-correlation property of the symbols in EPPM, when symbol $l$ is sent the expected value of $Z_l$, $E\{Z_l\} = K$, and $E\{Z_j\} = 0$ for $j \neq l$ [14]. So, in each symbol period, we form the set $\{Z_1, Z_2, \ldots, Z_Q\}$ and choose in favor of the largest element.

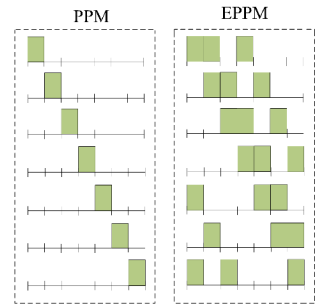

Fig. 1. Symbols for PPM and EPPM for $Q$=7.

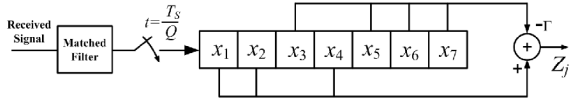

Fig. 2. Receiver for the EPPM code shown in Fig. 1. $T_s$ is the symbol period.

In AEPPM, when we include the complements of the BIBD codewords as symbols, we can use the same receiver structure as in Fig. 2. In that case, we extend the decision set to $\{Z_1, Z_2, \ldots, Z_Q, -Z_1, -Z_2, \ldots, -Z_Q\}$ and choose in favor of the largest element. The complexity of the decoder in this case is the same as for the EPPM decoder.

## III. PERFORMANCE ANALYSIS

We assume an additive white Gaussian noise channel with power spectral density $N_0/2$, as typical for RF systems and thermal or background noise limited FSO systems. We model the effect of this noise by adding a Gaussian random variable with variance $\Delta f N_0$ to the decision statistic $Z_j$, $j = 1, 2, \ldots, Q$, where $\Delta f$ is the receiver bandwidth. Since all symbols have equal energy, the optimum maximum likelihood (ML) decision rule reduces to the minimum distance criterion, and, therefore, the receiver in Fig. 2 is optimal. The union bound on the symbol error probability for an $M$-ary modulation can be expressed as [15]

$$P_s^{(U)} = \frac{1}{2M} \sum_{i=1}^{M} \sum_{\substack{j=1 \\ i \neq j}}^{M} \mathrm{erfc}\left(\sqrt{\frac{d_{ij}\gamma\eta}{2}}\right). \quad (6)$$

For an FSO system with bit-rate $R_b$, received peak optical power $P_0$ and photodetector responsivity $\rho$, the average SNR is $\gamma = \frac{\rho^2 P_0^2}{N_0 R_b}$ and the modulation efficiency is $\eta = \frac{\log_2 M}{Q}$. For an IR UWB system $\gamma = \frac{E_I R_b}{2\Delta f N_0}$, where $E_I$ is the impulse pulse energy, $\Delta f$ is independent from the bit-rate, and $\eta = \frac{Q}{\log_2 M}$.

For PPM, $M = Q$ and $d_{ij} = 2$, and so the union bound is

$$P_{s,\mathrm{PPM}}^{(U)} = \frac{Q-1}{2}\,\mathrm{erfc}\bigl(\sqrt{\gamma\eta}\bigr). \quad (7)$$

For MPPM, $M$ is $\binom{Q}{K}$ and for any symbol $j$, the number of symbols with distance $d$ to it is $\binom{K}{d/2}\binom{Q-K}{d/2}$. Thus, we have

$$P_{s,\mathrm{MPPM}}^{(U)} = \sum_{k=1}^{K-1} \binom{K}{k}\binom{Q-K}{k}\frac{1}{2}\,\mathrm{erfc}\bigl(\sqrt{k\,\gamma\eta}\bigr) \quad (8)$$

For high SNRs, the smallest distance between symbols limits $P_s$, so (8) can be approximated as

$$P_{s,\mathrm{MPPM}}^{(U)} \approx \frac{K(Q-K)}{2}\,\mathrm{erfc}\bigl(\sqrt{\gamma\eta}\bigr). \quad (9)$$

Similar to PPM, the union bound for EPPM is

$$P_{s,\mathrm{EPPM}}^{(U)} = \frac{Q-1}{2}\,\mathrm{erfc}\bigl(\sqrt{(K-\lambda)\gamma\eta}\bigr). \quad (10)$$

EPPM has a $(K - \lambda)$ coding gain advantage over PPM.

To calculate the BER, let $\boldsymbol{b}_\ell$, a $(\log_2 M)$-bit binary number, be the binary sequence assigned to symbol $\ell$. When the received symbol $\ell$ is estimated incorrectly as symbol $\ell'$, $d(\boldsymbol{b}_\ell, \boldsymbol{b}_{\ell'})$ bits are decoded incorrectly. Thus, for a $M$-ary modulation scheme, the BER can be upper bounded as [15]

$$P_b^{(U)} = \frac{1}{2M} \sum_{\ell=1}^{M} \sum_{\substack{\ell'=1 \\ \ell' \neq \ell}}^{M} \mathrm{erfc}\left(\sqrt{\frac{d_{\ell\ell'}\gamma\eta}{2}}\right)\frac{d(\boldsymbol{b}_\ell, \boldsymbol{b}_{\ell'})}{\log_2 M}. \quad (11)$$

For PPM and EPPM, since all symbol pairs are equidistant, the BER is independent of the bit-symbol mapping. Hence, according to (7) and (10), the BER for these two schemes reduces to $P_b = \frac{M}{2(M-1)} P_s^{(U)}$ [15, p. 399]. For AEPPM, we use the same expression as the worst case [15, p. 397]. For MPPM, unlike PPM and EPPM, the BER depends on the bit-symbol mapping. For larger constellation sizes finding the optimum bit-symbol mapping is a difficult problem for MPPM. Here, we use $P_b = P_s^{(U)}/\log_2 M$ as the best case BER for MPPM.

## IV. NUMERICAL RESULTS

In this section, numerical results are presented to compare the performance of EPPM with PPM and MPPM schemes for a constant bit-rate. The BER of 8-ary PPM ($Q = 8$), 64-ary MPPM ($Q = 12$, $K = 2$), 8-ary EPPM ($Q = 11$, $K = 5$), and 16-ary AEPPM ($Q = 11$, $K = 5$) are compared in Fig. 3 for an FSO link. The parameters are chosen so that the width of the time-slots for all modulation schemes are approximately equal. The simulations are done using Monte-Carlo method, and for each point at least 10/BER trials are run. The simulation results for MPPM are obtained using the optimal bit-symbol mapping presented in [16]. For all cases, the simulation results match well with the union bound for high SNRs. According to these results, EPPM requires a $\gamma$ 3.5 dB and 2.3 dB lower than PPM and MPPM, respectively, for a BER=$10^{-9}$. For AEPPM, since the constellation size is doubled, the BER is improved compared to EPPM.

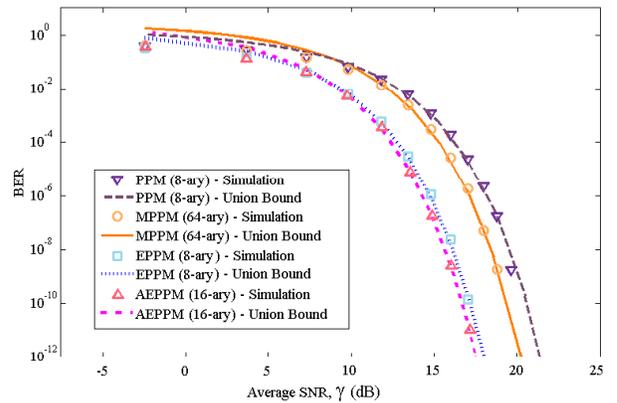

Fig. 3. BER of an FSO link versus $\gamma$ for 8-ary PPM, 64-ary MPPM, 8-ary EPPM, and 16-ary AEPPM.

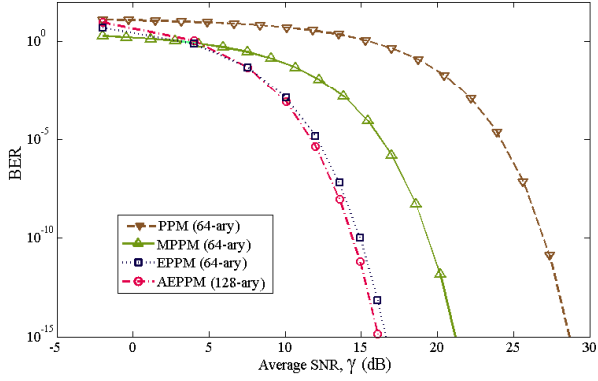

Fig. 4. Upper bound to BER versus $\gamma$ for 64-ary PPM, 64-ary MPPM, 64-ary EPPM, and 128-ary AEPPM.

In Fig. 4 the BERs for equal receiver complexity (number of correlations computed) are plotted using the union bound for an FSO system. For this case, the performance of the 64-ary EPPM and 128-ary AEPPM are considerably better than PPM and MPPM. EPPM requires a $\gamma$ 11.7 dB and 4.5 dB lower than PPM and MPPM, respectively, for a BER of $10^{-9}$. The symbol rate is the same for all modulation schemes, except for the AEPPM.

Fig. 5 shows the spectral efficiency, which is defined as the ratio of the bit-rate to the required receiver bandwidth, versus the required average SNR, $\gamma$, for a BER of $10^{-5}$, for PPM, EPPM, AEPPM and MPPM for FSO systems. Each point represents a scheme with different parameters. For PPM we include $Q = 2^i$ for $i = 2, 3, \ldots, 8$, and for EPPM results are obtained for $Q = 7, 11, 19, 35, 67, 131$ and $263$. For OOK, PPM, EPPM and AEPPM, by increasing $Q$ the spectral efficiency decreases. MPPM is able to achieve higher spectral efficiency since the constellation size is larger compared to other schemes. From these plots, PPM requires higher received peak power for larger $Q$, while the required $\gamma$ for EPPM and AEPPM decreases as $Q$ increases. Although MPPM can provide higher spectral efficiency, it needs a higher $\gamma$ for a fixed BER compared to EPPM. For the same $Q$, AEPPM provides higher spectral efficiency and requires lower received peak power compared to EPPM. In IR UWB systems, since the receiver bandwidth is determined by the pulse-width of the impulse, for a fixed bit-rate the spectral efficiency of all schemes will be equal.

## V. Conclusion

In this letter, a novel modulation scheme, called expurgated PPM, using the symmetric BIBD is introduced, and its advantages over PPM and MPPM schemes are described. Our new proposed approach is efficient for peak power limited communication systems and, therefore, can be used in FSO links and IR UWB communications. The bit-symbol mapping for EPPM is similar to PPM and can be easily implemented since all symbols are pair-wise equidistant. The symbols in EPPM can be cyclic shifts of each other, which simplifies the transmitter and receiver structure. In future work, we will consider optimal codes for dispersive peak-power limited channels.

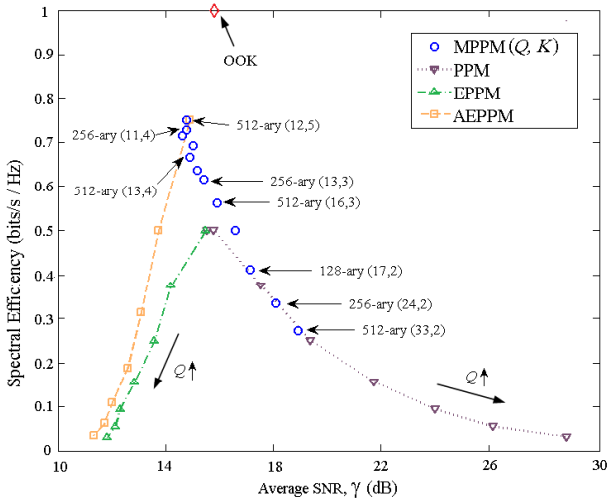

Fig. 5. Spectral efficiency and required $\gamma$ for BER of $10^{-5}$, for PPM, EPPM, AEPPM, MPPM and OOK.